\documentclass[reprint,aps,prd,nofootinbib]{revtex4-1}

\usepackage{amsmath}
\allowdisplaybreaks[4]
\usepackage{amsfonts}
\usepackage{xcolor}
\usepackage{graphicx}
\usepackage{epstopdf}
\usepackage{siunitx}

\usepackage{float}
\usepackage[
    pdfauthor={Jan-Simon Hennig},
    pdftitle={Demonstration of a switchable damping system to allow low-noise operation of high-Q low-mass suspended systems},
    colorlinks=true,
    linkcolor=black,
    urlcolor=blue,
    citecolor=blue
]{hyperref}

\DeclareSIUnit{\sqrthz}{\ensuremath{\sqrt{\text{\hertz}}}}

\begin{document}
\allowdisplaybreaks[4]
\title{Demonstration of a switchable damping system to allow low-noise operation of high-Q low-mass suspension systems}

\author{Jan-Simon Hennig$^1$}
\author{Bryan W. Barr$^1$}
\author{Angus S. Bell$^1$}
\author{William Cunningham$^1$}
\author{Stefan L. Danilishin$^{1,3,4}$}
\author{Peter Dupej$^1$}
\author{Christian Gr\"af$^1$}
\author{Sabina H. Huttner$^1$}
\author{Russell Jones$^1$}
\author{Sean S. Leavey$^1$}
\author{Daniela Pascucci$^1$}
\author{Martin Sinclair$^1$}
\author{Borja Sorazu$^1$}
\author{Andrew Spencer$^1$}
\author{Sebastian Steinlechner$^{1,2}$}
\author{Kenneth A. Strain$^1$}
\author{Jennifer Wright$^1$}
\author{Teng Zhang$^1$}
\author{Stefan Hild$^1$}
\affiliation{$1$ SUPA, School of Physics and Astronomy, The University 
of Glasgow, Glasgow, G12\,8QQ, UK}
\affiliation{$2$ Institut f\"ur Laserphysik und Zentrum f\"ur Optische Quantentechnologien der Universit\"at Hamburg, Luruper Chaussee 149, 22761 Hamburg, Germany}
\affiliation{$3$ Institut f\"ur Gravitationsphysik, Leibniz Universit\"at Hannover and Max-Planck-Institut f\"ur Gravitationsphysik (Albert-Einstein-Institut) Callinstr. 38, 30167 Hannover, Germany}
\affiliation{$4$ Institut f\"ur Theoretische Physik, Leibniz Universit\"at Hannover and Max-Planck-Institut f\"ur Gravitationsphysik (Albert-Einstein-Institut), Callinstr. 38, 30167 Hannover, Germany}

\begin{abstract}
Low mass suspension systems with high-Q pendulum stages are used to enable quantum radiation pressure noise limited experiments. Utilising multiple pendulum stages with vertical blade springs and materials with high quality factors provides attenuation of seismic and thermal noise, however damping of these high-Q pendulum systems in multiple degrees of freedom is essential for practical implementation. Viscous damping such as eddy-current damping can be employed but introduces displacement noise from force noise due to thermal fluctuations in the damping system. In this paper we demonstrate a passive damping system with adjustable damping strength as a solution for this problem that can be used for low mass suspension systems without adding additional displacement noise in science mode. We show a reduction of the damping factor by a factor of 8 on a test suspension and provide a general optimisation for this system. 
\end{abstract}

\maketitle


\section{\label{sec:introduction}Introduction}
The recent discoveries of gravitational waves (GW) \cite{Abbott2016,Abbott2016a,Abbott17} have firmly established a new era of gravitational wave astronomy. To achieve the required sensitivity GW detectors such as Advanced LIGO (aLIGO) \cite{Aasi2015} employ several techniques to reduce unwanted noise sources. One limiting noise source at low frequencies is seismic noise. To attenuate the influence of this limiting noise source the mirrors are suspended as cascaded multi-stage pendulums \cite{Aston2012}, in this paper referred to as suspensions, and are additionally mounted on actively controlled seismic attenuation tables \cite{Matichard2015}. 

In this configuration every pendulum stage acts as a harmonic oscillator with a $1/f^2$ response above the highest pendulum mode-frequency attenuating any noise introduced at the top stage of the pendulum. A quadruple pendulum, as used in aLIGO, has a $1/f^8$ response above the highest pendulum mode-frequency. However, the motion at these mode-frequencies of the pendulum needs to be attenuated. To avoid limitation by thermal noise the use of materials with high quality factors (high-Q) is essential. This is one reason the lowest stage of the aLIGO test mass suspension is fully monolithic using fused silica masses and fibres of the same material \cite{Cumming2012}. Creating a high-Q system like this means that mode-frequencies, once excited, ring down only after very long times, as 
\begin{equation}
	A(t)=A(0)\exp{\left[-\frac{\omega_{\rm{res}}}{2Q} t\right]},
\end{equation}
where $A(0)$ denotes an arbitrary initial amplitude, $t$ is time and $\omega_{\rm{res}}$ denotes the angular resonance frequency of the excited mode.

There are two ways of damping these resonances, one using active damping \cite{Plissi2000} the other using passive damping \cite{Plissi2004}. The method described in ref.~\cite{Plissi2000} uses active damping employing a coil-magnet actuator and a sensor. The sensors measure the motion of the pendulum (shadow sensors and optical levers are used in the case of aLIGO \cite{Carbone2012}) and create a feedback signals for the coil-magnet actuator. By sending a current through the coil, a counteracting force can be applied to the pendulum to damp its motion. Passive damping is described in ref.~\cite{Plissi2004} where a magnet and conductive surface are arranged such that when the magnet moves relative to the conductive material eddy-currents are induced which creates Joule heating due to the residual resistance of the material; this heating represents power loss and thus damping of the system. In this case the damping is viscous and thus velocity dependent.

In Section~\ref{sec:PDvsTN} we show that thermal noise of a coil-magnet actuator and required damping for a low mass pendulum system as used in the Glasgow Sagnac speed meter \cite{Graf2014} are two contradicting goals. Section~\ref{sec:Switchable} presents a solution to this problem and discusses the results of a switchable eddy-current damping system. In Section~\ref{sec:optiR} we highlight a possible optimisation for separate mode-frequencies.


\section{\label{sec:PDvsTN}Damping and related force noise}
Passive eddy-current damping is used at the uppermost stage of the \SI{100}{\gram} triple suspensions for optical spring experiments at the \SI{10}{\metre} prototype in Glasgow \cite{Edgar2011}. A coil is wound around a copper former and a magnet is fixed onto the pendulum giving rise to passive damping, actuation on the pendulum and, in the presence of a sensor, could give rise to active damping. However, in this case another noise source becomes more dominant: due to the use of a conductive material such as copper and a coil-magnet actuator, the amplitude spectral density of force noise due to thermal noise of the coil-magnet actuator in [\si{\newton\per\sqrthz}] is given by \cite{Plissi2004}
\begin{equation}\label{eq:TN}
\tilde{f}=\sqrt{4k_\text{B}T\gamma},
\end{equation}
where $\gamma$ is the the eddy-current damping constant of the coil-magnet actuator, $k_\text{B}$ is the Boltzmann constant and $T$ is temperature. The displacement spectral density can be deduced from multiplication with the corresponding force transfer function of the pendulum. 

\begin{figure}
	\includegraphics[width=0.45\textwidth]{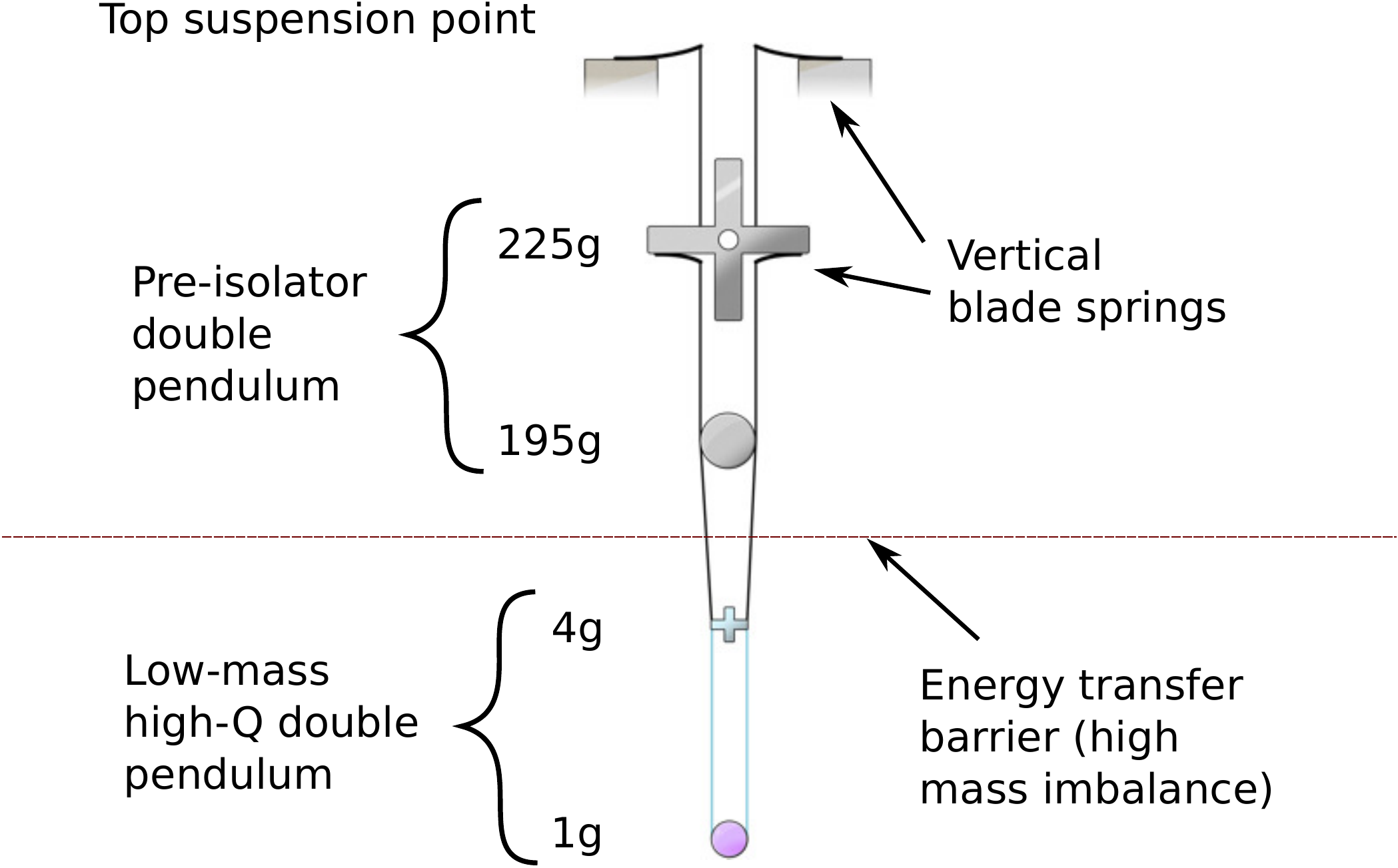}
	\caption{Schematic of the \SI{1}{\gram} ITM suspension. The small vertical blade springs dictate the use of a fourth pendulum stage above the \SI{1}{\gram} test mass to allow sufficient energy transfer for damping resonance mode-frequencies of the low-mass high-Q double pendulum.}
	\label{fig:1gscheme}
\end{figure}

From Equation~(\ref{eq:TN}) it is clear that high damping and low force noise cannot be achieved simultaneously. In general this is not a problem as passive damping is mostly used in the uppermost stage of the pendulum and thus filtered by the pendulum stages below. For certain other applications this can lead to an issue. The so-called Sagnac speed meter (SSM) experiment, currently being commissioned in Glasgow, uses \SI{1}{\gram} input test masses (ITMs) to ensure limitation by quantum radiation pressure noise in the desired operating band (hundreds of \si{\hertz}) \cite{Graf2014}. To minimise suspension thermal noise, the lowest stage is fully monolithic and employs fused silica fibres of diameter \SI{10}{\micro\metre}. This creates high-Q pendulum mode-frequencies of the lowest stage which need to be damped whenever they are excited, e.g. when the cavities fall out of lock. The mode-frequencies that are most likely being excited by this and need to be damped for practical implementation of these low mass suspension systems are longitudinal, rotation (yaw) and tilt (pitch). The ITM suspension is designed as a quadruple pendulum as shown in Figure~\ref{fig:1gscheme}.
For vertical isolation two stages of small blade springs are used at the suspension point and the top mass. The smallest blades at the top mass require a minimal load of \SI{100}{\gram} per blade. In order to extract sufficient energy from the two orders of magnitude lighter ITMs an extra pendulum stage must be introduced to the design above the test mass, creating a two-stage heavy pre-isolator pendulum from which a low-mass high-Q double pendulum is suspended. 

\begin{figure}
	\includegraphics[width=0.5\textwidth]{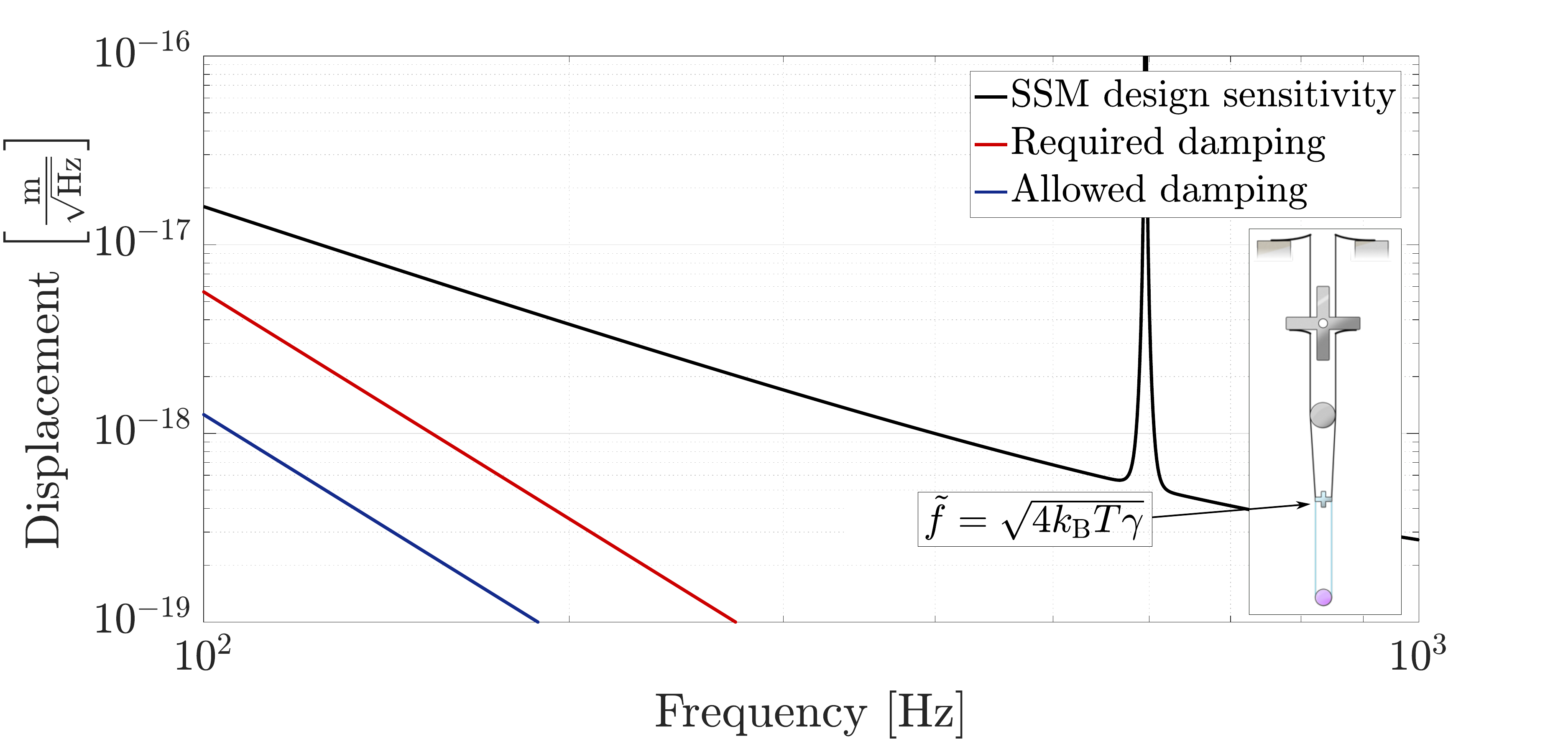}
	\caption{Comparison of allowable to required damping for a \SI{1}{\gram} suspension used in the SSM. The required damping is derived from ring-down times of the longitudinal mode-frequency where times above \SI{10}{\second} are considered impractical. The allowable damping considers a safety factor of 10 below design sensitivity for one double pendulum.}
	\label{fig:DvsTN}
\end{figure}
Figure~\ref{fig:DvsTN} shows the simulated displacement spectral density for a double pendulum with a \SI{1}{\gram} test mass and a \SI{4}{\gram} penultimate mass due to force noise of a coil-magnet actuator acting on the penultimate stage for two damping constants. This figure is based on the longitudinal force transfer function of the pendulum given in [\si{\metre\per\newton}]. The case for required damping is depicted in the red curve. With a high damping constant we can achieve a simulated ring-down of \SI{10}{\second} for the longitudinal resonance frequency. However, this case does not fulfil the desired safety factor of times 10 in displacement sensitivity at \SI{100}{\hertz}. The case for allowable damping, shown in blue, fulfils the safety factor, but results in impractical ring-down times of order 100s of seconds due to a low damping constant. The opposing goals of required damping and low displacement noise cannot be achieved simultaneously unless the required damping can be switched to a significantly lower value during science mode of the experiment. 


\section{\label{sec:Switchable}Measurement of Switchable ECD}
\begin{figure}[hbt]
	\includegraphics[width=0.45\textwidth]{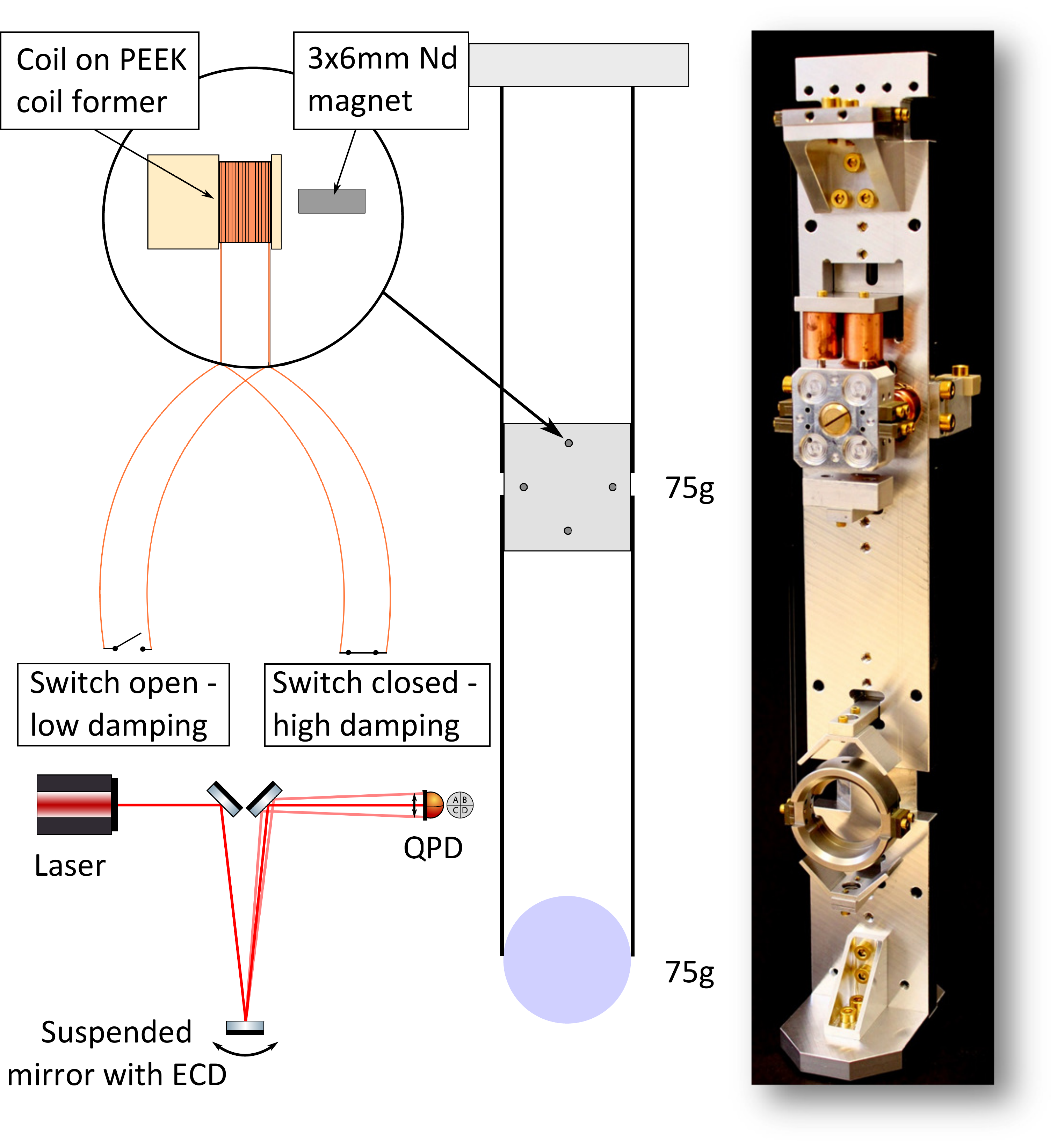}
	\caption{Experimental realisation of switchable ECD. Top left: The coils are wound around non-inductive PEEK coil formers. Short-circuiting the coils results in damping (dependent on residual resistance of coil) and opening the coils and attaching one end to ground (current drive) or adding a big resistance between the coil ends (voltage drive) results in low damping but allows for actuation. Bottom left: Measurements are performed in an optical lever set up. Right: Picture of the suspension.}
	\label{fig:setup}
\end{figure}

\begin{figure}[hbt]
	\includegraphics[width=0.5\textwidth]{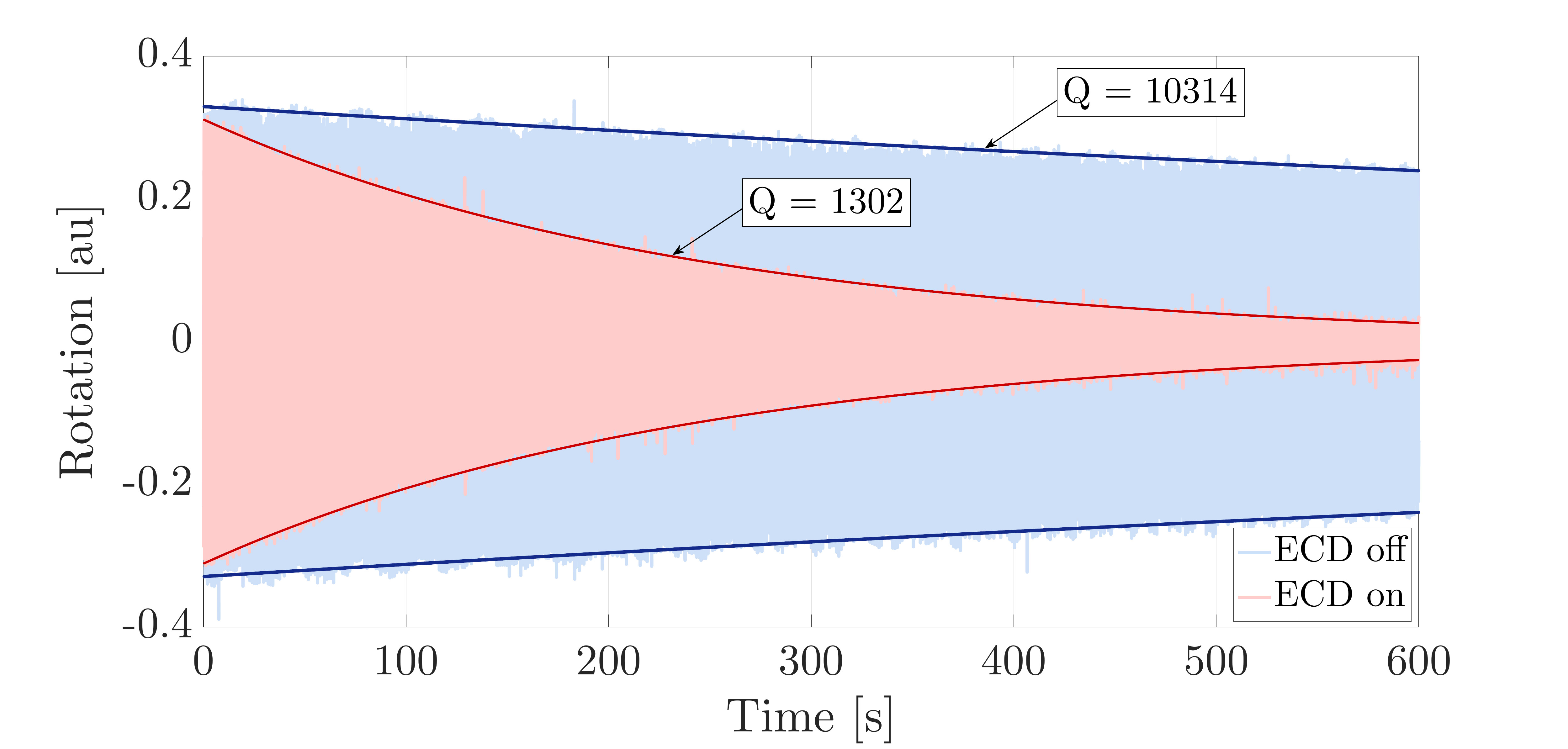}
	\caption{Measurement of switchable ECD on a test suspension in yaw. Light blue shows the ring-down measurement in ECD off (switches open) mode and light red shows the ring-down measurement in ECD on mode. The solid blue and solid red curves are fitted to the maxima of the corresponding ring-down measurement. We achieve a reduction of the Q-factor of the yaw mode of a factor of 8 in this test experiment.}
	\label{fig:result}
\end{figure}

A switchable eddy-current damper can be implemented with modified standard coil-magnet actuators. The damping of this system depends on the response of the coil-magnet actuator and the residual resistance of the coil. The higher the residual resistance in the coil, the lower the damping of the system:
\begin{equation}
\gamma \approx \frac{V^2}{R}, \label{eq:damping}
\end{equation}
where $\gamma$ denotes the damping factor in [\si{\kilogram\per\second}], $V$ the response of the coil-magnet actuator in [\si{\newton\per\ampere}] and $R$ the residual resistance in [\si{\ohm}]\footnote{This approximation is frequency independent. We will discuss frequency effects later on in Section~\ref{sec:optiR}.}.
Equation~\ref{eq:damping} shows that breaking the coil loop or using a high resistance will result in low or negligible damping. In order to demonstrate the reduction in Q of a pendulum resonance mode-frequency by switching the eddy-current damper on and off we use a double pendulum suspension with masses of \SI{75}{\gram} for the top and test mass. A picture of this type of suspension is shown in Figure~\ref{fig:setup} (right).
Instead of conductive copper coil formers, which would provide additional damping, Polyether ether ketone (PEEK) formers are used. We use four coils at the penultimate pendulum stage mounted rigidly to the suspension frame and four 3x\SI{6}{\milli\metre} nickel coated neodymium magnets glued to the penultimate mass. The coils are formed by \SI{200}{\micro\metre} enamelled copper wire with 150 turns leading to a residual resistance of \SI{3.6}{\ohm} as measured at the switch terminals. The coil-magnet actuators are arranged in a cross configuration through the centre of mass of the top mass which allows damping of yaw, pitch and longitudinal mode. The schematic layout can be seen in Figure~\ref{fig:setup} (top left). The measurement is performed using an optical lever (see Figure~\ref{fig:setup} bottom left). We use a silver-coated mirror as a test mass in the suspension and a quadrant photodiode (QPD) for the readout of ring-down times for the lower frequency yaw mode which we use as an illustrative example in this paper. The lower pendulum stage of a double pendulum has two yaw modes, one common (lower frequency) and one differential (higher frequency) mode between the two masses\footnote{To show a clear and clean ring-down measurement we use only one coil for damping of this system.}. For a measurement of the ring-down of a specific resonance mode-frequency the coil-magnet actuators are used to excite the pendulum with an impulse. Shortly after, depending on which case is measured, the switch to short circuit the coil is closed (ECD on) or remains open (ECD off). The data from the QPD is recorded with the Control and Data System (CDS) \cite{Bork2010} and analysed in \textsc{MATLAB} using its internal fitting toolbox. We use the maxima of the oscillation for an exponential fit and calculate Q-factor and damping factor. The measurement data and fitted curve for the yaw mode can be seen in Figure~\ref{fig:result}.
This measurement was taken over 600 seconds and the improvement in damping can be quantified by the calculated Q-factor of about 10300 for this mode when the ECD is off and a Q-factor of about 1300 when the ECD is on. The damping factor for the ECD switched on can thus be calculated to be \SI{1.26}{\gram\per\second} and \SI{0.16}{\gram\per\second} for the ECD switched off. The difference between these values, \SI{1.10}{\gram\per\second}, is the damping factor that can be attributed to the coil-magnet actuator. We assume one coil acts as a damper for this rotational measurement. The coil-magnet actuator was calculated to have a response of \SI{0.105}{\newton\per\ampere}. Together with the measured residual resistance of the coil of \SI{3.6}{\ohm} the damping for this coil can be deduced to be \SI{3.0625}{\gram\per\second}. The mismatch between measured and calculated damping factor can be explained by alignment of the magnet relative to the coil. In this case the coil former was intended to be used with smaller 2x\SI{2}{\milli\metre} magnets and thus does not allow the bigger 3x\SI{6}{\milli\metre} magnets to be placed in the actuation sweet spot of the coil but \SI{2.5}{\milli\metre} offset resulting in lower response of the actuator. Our measurement indicates a reduction in the damping factor of one coil by a factor of 8 when the ECD is switched off (science mode). Depending on which coils are used in the local control assembly, switchable ECD can as well be applied for pitch and longitudinal mode with the same reduction in damping factor per coil. 

For the \SI{1}{\gram} ITM suspension of the Glasgow Sagnac speed meter experiment it is planned to integrate four coils with switchable ECD at the penultimate mass. In this design each coil will contribute a damping factor of \SI{2.4}{\gram\per\second} and ring-down times of certain resonance mode-frequencies will be reduced to $\sim$\SI{10}{\second}. The total damping is a factor of two bigger for the longitudinal mode (four coils) than for the rotational modes yaw and pitch (two coils). 


\section{Optimisation of Damping}
\label{sec:optiR}
\begin{figure}[hbt]
	\includegraphics[width=0.5\textwidth]{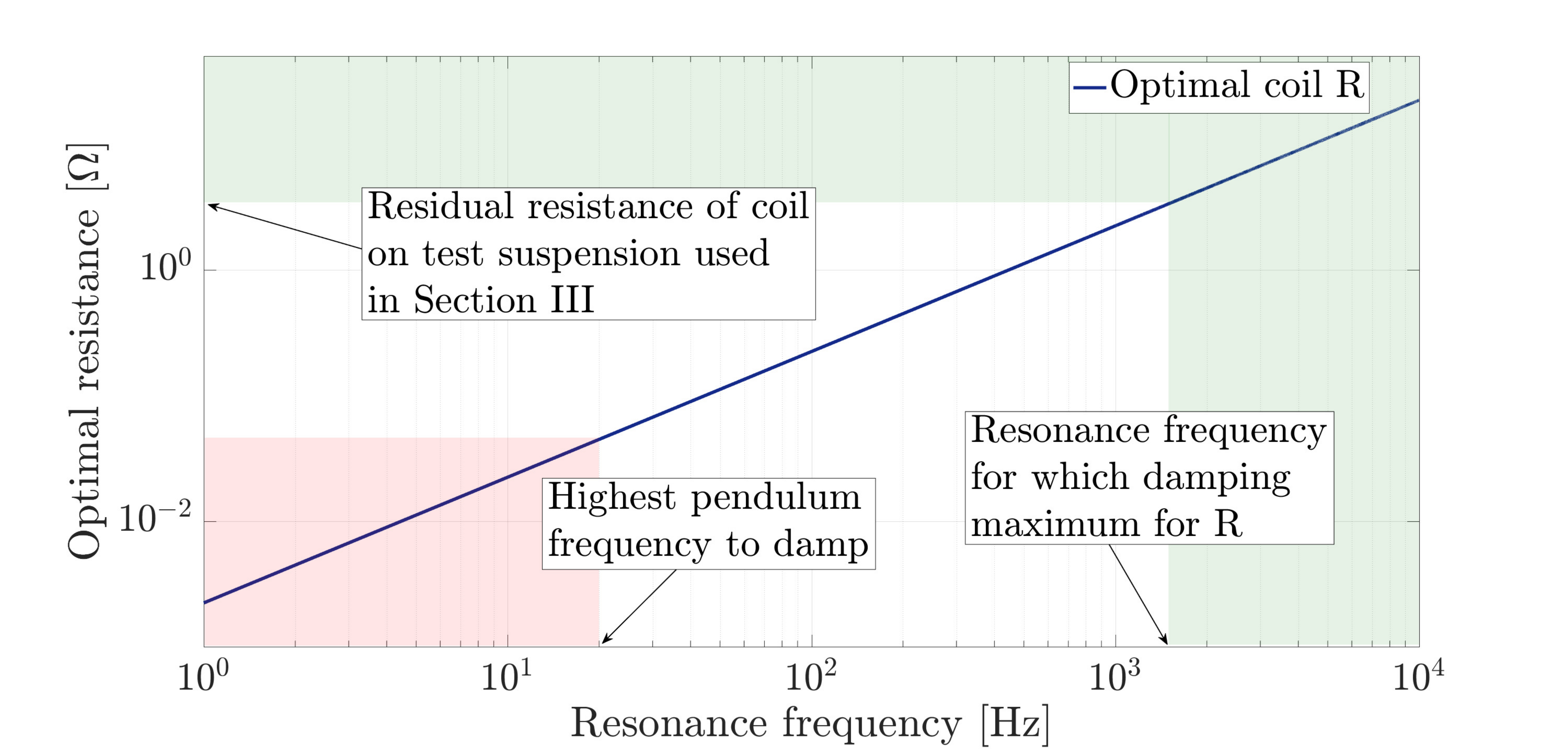}
	\caption{Optimal residual coil resistance in dependence of resonant mode-frequency assuming a certain coil design with an inductance of \SI{355}{\micro\henry}. The red shaded area represents the area in which resonant frequencies of a pendulum suspension typically lie. The optimal resistance is far below the \SI{3.6}{\ohm} coil resistance. The green shaded area shows the area above the coil resistance and thus the area in which optimisations could be achieved.}
	\label{fig:optimalR}
\end{figure}

In the previous section we used a frequency independent approximation for the damping factor. This approximation is only valid for small values of coil inductance $L$ and/or resonance mode-frequencies $\omega_{\rm{res}}$. The damping factor can in general be optimised for coil resistance. The short-circuited coil can be described as a RL circuit. The motion of a magnet inside a coil generates an electromotive force (EMF) and hence a current in the coil. Due to this magnetic induction a counter EMF is generated at the same time which opposes the initial flow of current. The flow of current and thus damping of the coil-magnet actuator can be optimised by impedance matching of any additional resistance to the RL-circuit. Equation~\ref{eq:damping} suggests to aim for lowest resistance, but it can be shown that this approximation is only valid for low inductances of the coil and/or low resonance frequencies. For a given mode-frequency $\omega_{\rm{res}}$ and given inductance $L$ the optimal impedance matched value of the resistance $R$  for a given RL-circuit can be calculated from 
\begin{equation}
\omega_{\rm{res}} = \frac{R}{L}
\end{equation}
The coils used in this experimental set-up described in Section~\ref{sec:Switchable} have a calculated inductance\footnote{With the relative permeability of a neodymium magnet of 1.05 the upper bound for the coil inductance is \SI{372.75}{\micro\henry}.} of \SI{355}{\micro\henry}. Figure~\ref{fig:optimalR} shows the optimal coil resistance $R$ in dependence of resonance mode-frequency. 
Considering mode-frequencies for the double pendulum used below \SI{20}{\hertz}, the optimal values for $R$ are below \SI{450}{\milli\ohm} and thus below the residual resistance of the coil wire itself (\SI{3.6}{\ohm}). From this figure we can deduce that there is indeed an optimal value for $R$ for given coil inductance $L$ and that for higher mode-frequencies this becomes an important optimisation to consider. However, for the pendulum mode-frequencies as well as coil dimensions in this paper this optimisation is not feasible. 


\section{Conclusion}
In this paper we have demonstrated an adjustable passive damping system with switchable eddy-current dampers as a solution for the problem of force noise arising from thermal fluctuations of viscous damping in a high-Q pendulum system. High damping for practical ring-down times of the pendulums mode-frequencies and low thermal noise due to damping in a coil-magnet actuator are two opposing goals in suspension design. In our system our measurements showed that switchable eddy-current damping allows for a reduction in the damping factor of the chosen coil-magnet actuator by a factor of 8 for one coil and thus reduces force noise due to thermal noise of the actuator. Depending on the coil-magnet actuator design varying factors are possible to achieve.

A general optimisation of the damping for certain mode-frequencies is possible by choosing the impedance matched value for the residual resistance. For the actuator design presented in this paper, this optimisation is not possible due to the high residual resistance of the coil wire and the low mode-frequencies of the double pendulum.

\section{Acknowledgements}
We want to thank Brett Shapiro for his valuable comments to improve the quality of this manuscript. The work described in this article is funded by the European Research Council (ERC-2012-StG: 307245). We are grateful for support from Science and Technology Facilities Council (Grant Ref: ST/L000946/1), the Humboldt Foundation, the International Max Planck Partnership (IMPP) and the ASPERA ET-R\&D project. This paper has LIGO document No. P1700333.


\bibliography{SwitchableECD_arxiv}

\end{document}